\documentclass[reqno]{amsart}
\usepackage{amssymb}
\usepackage{upref}
\begin{document}
\title[Transformations of ordinary differential equations]{Transformations of
ordinary differential equations via Darboux transformation technique}
\author{N.V. Ustinov}
\thanks{Supported by RFFI grant \symbol{242} 97--01--00752}
\address{Department of Theoretical Physics, Kaliningrad State University,
Al. Nevsky street 14, 236041, Kaliningrad, Russia}
\begin{abstract}
A new approach for obtaining the transformations of solutions of nonlinear
ordinary differential equations representable as the compatibility condition
of the overdetermined linear systems is proposed.
The corresponding transformations of the solutions of the overdetermined
linear systems are derived in the frameworks of the Darboux transformation
technique.
\end{abstract}
\maketitle
\section{Introduction}
The nonlinear ordinary differential equations (ODE's) admitting a
representation as the compatibility condition of the overdetermined system of
linear equations (Lax pair) are intensively studied in modern mathematical
physics.
The most famous equations belonging to this class are the six
Painlev$\acute{\mbox{\rm e}}$ equations $\rm P_I$--$\rm P_{VI}$ \cite{1,2,3}.
It was found that these equations are closely connected with nonlinear
partial differential equations (PDE's) integrable in the frameworks of the
inverse scattering transformation (IST) method \cite{4}.
Various approaches developed in the theory of nonlinear integrable PDE's
were applied to study ODE's of such a class (e.g., see reviews \cite{5,6} and
references therein).

It is well known that $\rm P_{II}$--$\rm P_{VI}$ have the transformations that
map the solutions of a given Painlev$\acute {\mbox{\rm e}}$ equation into
solutions of the same equation but with different values of the parameters of
the equation.
Such transformations for $\rm P_{II}$ were found in \cite{7} as a
generalization of corresponding formulas of the rational solutions
transformations \cite{8}.
The transformations of solutions of $\rm P_{II}$ were also derived by means of
the B\"acklund transformation technique \cite{9}.
A representation of the Painlev$\acute {\mbox{\rm e}}$ equations as the systems
of the first order equations was used to obtain the transformations of
$\rm P_{III}$ in the cases of two different choices of values of the parameters
\cite{10,11}, $\rm P_{IV}$ \cite{12} and of $\rm P_V$ with special parameter
values \cite{11}.
In similar manner the transformations for $\rm P_{II}$ and $\rm P_{IV}$ were
obtained in \cite{13}.
Moreover, an equivalence of nonlinear ODE to a system of equations was used to
construct the transformations of solutions of the higher order nonlinear ODE's
appearing as the self--similar reduction in the KdV hierarchy \cite{13,14}.
An unified approach to investigate the transformation properties of
Painlev$\acute {\mbox{\rm e}}$ equations, which utilizes the singularity
structure of their solutions, were developed in \cite{15}.
The known transformations and new ones for $\rm P_V$, that differs from the
equation considered in \cite{11} by a region of the parameters values, and
for $\rm P_{VI}$ were derived.
The truncation method was modified to find the transformations of nonlinear
ODE's \cite{16}.
Its extension, which is based on considering the transformations that preserve
the locations of subsets of the singularities of solutions, was suggested
\cite{17}.
The transformations of solutions of $\rm P_{II}$ and $\rm P_{IV}$ were, in
particular, obtained using this approach.

The transformations of nonlinear ODE representable as the compatibility
condition of the Lax pair cause evidently the transformations of the Lax pair
solutions.
Studying the monodromy preserving deformations of matrix linear ODE's with
regular singularities, Schlesinger \cite{18} constructed, in pure algebraic
way, the transformations that keep the monodromy data of the solutions except
for the exponents of formal monodromy.
These transformations, which are now referred to as the Schlesinger
transformations, were generalized for linear equations with irregular singular
points by Jimbo and Miwa \cite{3}.
The monodromy preserving condition is equivalent to having a set of additional
linear ODE's on solution of the isomonodromic equation \cite{18,19}, which
establish a hierarchy of the Lax pairs.
However, the authors mentioned seem to apply no the Schlesinger
transformations for the derivation of transformations for nonlinear ODE's
(such as Painlev$\acute{\mbox{\rm e}}$ equations; see \cite{3}, p. 437) that
admit the compatibility condition representation.
The transformations of $\rm P_{II}$--$\rm P_{VI}$, including new ones, were
obtained by means of the Schlesinger transformations of associated Lax pairs
in \cite{20}.

In the present paper we show that the Darboux transformation (DT) technique
\cite{21} can be used to construct the transformations of solutions of
nonlinear ODE's and associated Lax pairs.
This technique is suitable for generating the infinite hierarchies of
solutions satisfying the reduction constraints imposed on the coefficients of
the Lax pairs.
To extend the DT technique for obtaining the transformations of ODE's we
suggest that the points, in which the DT possesses a kernel, coincide with the
singular points of the Lax pairs.
The cases of $\rm P_{II}$ and certain nonlinear ODE of the second order are
examined by way of illustration.

The paper is organized as follows.
The Lax pairs of the ODE's under consideration are written in Sec.2.
We present the explicit formulas of the DT's in Sec.3 and discuss the
properties of them.
The asymptotic expansions at the singular point neighborhood of the Lax pairs
solutions are given in Sec.4.
These expansions are used there to build the transformations of the ODE's on
the basis of the DT formulas.

\section{Lax Pairs and Nonlinear ODE's}
Let us consider direct Lax pair
\begin{equation}
\left\{
\begin {array}{l}
\Psi_x = P \Psi\\
\Psi_\lambda = Q \Psi\\
\end {array}
\right.
\label{1}
\end{equation}
for matrix function $\Psi \equiv \Psi (x, \lambda)$ and dual Lax pair
\begin{equation}
\left\{
\begin{array}{l}
\Xi_x = -\Xi P\\
\Xi_{\lambda} = -\Xi Q\\
\end{array}
\right.
\label{2}
\end{equation}
for matrix $\Xi \equiv \Xi (x, \lambda)$.
The compatibility condition  of the direct pair
$\Psi_{x \lambda} = \Psi_{\lambda x}$ (or, for dual one,
$\Xi_{x \lambda} = \Xi_{\lambda x}$) leads
to the condition on the matrix coefficients $P$ and $Q$:
\begin{equation}
P_\lambda - Q_x + [P,Q] = 0 .
\label{3}
\end{equation}

The equation $\rm P_{II}$
\begin{equation}
u_{xx} = 2 u^3 + x u - \alpha
\label{4}
\end{equation}
arises from the compatibility condition of Lax pairs (\ref{1},\ref{2})
with coefficients
\begin{equation}
P = - \lambda \sigma_3 + u \sigma_1 ,
\label{5}
\end{equation}
\begin{equation}
Q = 4 \lambda^2 \sigma_3 -
4 \lambda u \sigma_1 + \sigma_3 \left(2 u_x \sigma_1 -
2 u^2 E - x E\right) + \lambda^{-1} \alpha \sigma_1 ,
\label{6}
\end{equation}
where $\sigma_i$ ($i=1,2,3$) are the Pauli spin matrices, $E$ is $2*2$ unit
matrix.

In the sequel the DT technique will be applied for obtaining the
transformations of the solutions of $\rm P_{II}$ and following ODE
\begin{equation}
x^2 u u_{xx} = (x u_x)^2 - x u u_x + \alpha^2 u^4 - 2 \gamma  u^3 -
2 ( 2 \beta + 1) x u - 4 x^2
\label{7}
\end{equation}
($\alpha$, $\beta$ and $\gamma$ are constants), for which the transformations
were not written explicitly in the literature.
This equation is connected with special cases of $\rm P_{III}$ and
$\rm P_V$ \cite{1,22}.

Let us consider direct and dual Lax pairs (1,2) with coefficients
\begin{equation}
P = - \lambda \sigma_3	+ \displaystyle \frac{1}{x}
\left(
\begin{array}{cc}
0 & v_1 \\
v_2 & 0
\end{array}
\right) ,
\label{8}
\end{equation}
\begin{equation}
Q = - x \sigma_3 + \lambda^{-1}
\left(
\begin{array}{cc}
\beta & v_1 \\
v_2 & - \beta
\end{array}
\right)
+ \lambda^{-2}
\left(
\begin{array}{cc}
q & w_1 \\
w_2 & - q
\end{array}
\right) ,
\label{9}
\end{equation}
where
$
v_1 = u z w , \,\,\,
v_2 = \displaystyle \frac{v (\alpha - z)}{w} , \,\,\,
w_1 = z w , \,\,\,
w_2 = \displaystyle \frac{\alpha - z}{w} , \,\,\,
q = z - \displaystyle \frac{\alpha}{2} .
$
The compatibility condition (\ref{3}) in this case yields
$$
\left\{
\begin{array}{l}
u_x = - 2 - \frac{\displaystyle 2 \beta}{\displaystyle x} u +
\frac{\displaystyle 2 z - \alpha}{\displaystyle x} u^2 , \\
v_x = 2 + \frac{\displaystyle 2 \beta}{\displaystyle x} v -
\frac{\displaystyle 2 z - \alpha}{\displaystyle x} v^2 , \\
z_x = \frac{\displaystyle v - u}{\displaystyle x} z (z - \alpha) , \\
\frac{\displaystyle w_x}{\displaystyle w} = -
\frac{\displaystyle u z + v ( z - \alpha)}{\displaystyle x} .
\end{array}
\right.
$$
The first three equations are reduced to Eq.(\ref{7}).

\section{Darboux Transformation Technique}
The first equations of Lax pairs (\ref{1},\ref{2}) with coefficient $P$
given by Eq.(\ref{5}) or Eq.(\ref{8}) are well known in the theory of IST
as direct and dual Zakharov--Shabat (ZS) spectral problems
\begin{equation}
\Psi_x = (\lambda P^{(1)} + P^{(0)}) \Psi ,
\label{10}
\end{equation}
\begin{equation}
\Xi_x = -\Xi (\lambda P^{(1)} + P^{(0)}) .
\label{11}
\end{equation}
Potential $P^{(0)}$ of these problems has following general form:
$$
P^{(0)} =
\left(
\begin{array}{cc}
0 & u_1 \\
u_2 & 0
\end{array}
\right) .
$$

The DT technique allows one to produce new solutions of the ZS problems and
corresponding matrix potential, starting from the initial ones.
Let $\varphi=(\varphi_1,\varphi_2)^T$ be a vector solution of direct ZS
problem (\ref{10}) with $\lambda = \mu$.
The first elementary DT (EDT) of direct problem
$\{\Psi,\Xi,P^{(0)}\}\to\{\tilde\Psi,\tilde\Xi,\tilde P^{(0)}\}$ is defined
as given \cite{23,24}:
\begin{equation}
\tilde \Psi = \sigma
\left(
\begin{array}{cc}
\lambda - \mu + \displaystyle \frac{u_1 \varphi_2}{2\varphi_1} &
- \displaystyle \frac{u_1}{2}  \\
- \displaystyle \frac{\varphi_2}{\varphi_1} & 1
\end{array}
\right)
\Psi , \,\,\,
\tilde \Xi = \displaystyle \frac{\Xi}{\sigma}
\left(
\begin{array}{cc}
1 & \displaystyle \frac{u_1}{2}  \\
\displaystyle \frac{\varphi_2}{\varphi_1} &
\displaystyle \lambda - \mu + \frac{u_1 \varphi_2}{2 \varphi_1}
\end{array}
\right),
\label{12}
\end{equation}
\begin{equation}
\tilde P^{(0)} =
\left(
\begin{array}{cc}
0 & \displaystyle - \frac{u_{1,x}}{2} + \frac{u_1^2 \varphi_2}{2 \varphi_1} -
\mu u_1 \\
\displaystyle 2 \frac{\varphi_2}{\varphi_1} & 0
\end{array}
\right)
\label{13}
\end{equation}
(it is assumed hereafter that $\sigma$ is a scalar function of $\lambda$).
The direct and dual ZS problems are covariant with respect to this EDT:
matrix functions $\tilde\Psi$ and $\tilde\Xi$ are the solutions of
Eqs.(\ref{10},\ref{11}) with potential $\tilde P^{(0)}$.
The second EDT of direct problem has a form:
\begin{equation}
\tilde \Psi = \sigma
\left(
\begin{array}{cc}
1 & - \displaystyle \frac{\varphi_1}{\varphi_2} \\
\displaystyle \frac{u_2}{2} &
\lambda - \mu - \displaystyle \frac{u_2 \varphi_1}{2 \varphi_2}
\end{array}
\right)
\Psi , \,\,\,
\tilde \Xi = \displaystyle \frac{\Xi}{\sigma}
\left(
\begin{array}{cc}
\lambda - \mu - \displaystyle \frac{u_2 \varphi_1}{2 \varphi_2} &
\displaystyle \frac{\varphi_1}{\varphi_2} \\
- \displaystyle \frac{u_2}{2} & 1
\end{array}
\right),
\label{14}
\end{equation}
\begin{equation}
\tilde P^{(0)} =
\left(
\begin{array}{cc}
0 & \displaystyle - 2 \frac{\varphi_1}{\varphi_2} \\
\displaystyle \frac{u_{2,x}}{2} - \frac{u_2^2 \varphi_1}{2 \varphi_2} -
\mu u_2 & 0
\end{array}
\right) .
\label{15}
\end{equation}

In similar manner one can define the EDT's of dual problem, which depend
explicitly on a solution of the dual ZS problem (\ref{11}).
Successive carrying out of an EDT of direct problem and proper one of
dual problem leads to so--called binary DT (BDT) \cite{25,26}:
\begin{equation}
\tilde \Psi = \sigma \left( E - \frac{\displaystyle\mu - \nu }
{\displaystyle\lambda - \nu } \, R \right) \Psi , \,\,\,
\tilde \Xi = \displaystyle \frac{\Xi}{\sigma} \left( E - \frac{\displaystyle\mu - \nu }
{\displaystyle\mu - \lambda} \, R \right) ,
\label{16}
\end{equation}
\begin{equation}
\tilde P^{(0)} = P^{(0)} + (\mu - \nu) \left[ P^{(1)} , R \right] ,
\label{17}
\end{equation}
where matrix R is the projector:
$
R_{ij} = \varphi_i \chi_j / (\chi, \varphi)
$,
$\varphi=(\varphi_1,\varphi_2)^T$ is the vector solution of Eq.(\ref{10}) with $\lambda = \mu$,
$\chi=(\chi_1,\chi_2)$ is the vector solution of Eq.(\ref{11}) with $\lambda = \nu$.

It is seen that the potential in the Lax pairs for $\rm P_{II}$ is the
symmetric matrix.
The DT technique is convenient for keeping the reduction constraints imposed
on the coefficients of the spectral problems \cite{27}.
We can obtain the transformed potentials to satisfy this reduction, performing the
iterations of BDT \cite{25,26}.

The DT's presented in this section allow us to build infinite hierarchy of
solutions of nonlinear PDE's integrable in the frameworks of the IST method.
It is supposed in so doing that vectors $\varphi$ and $\chi$ are the solutions
of Lax pairs associated with given PDE.
Unfortunately, this approach is unfit for obtaining the transformations, which
generate infinite hierarchy of solutions of nonlinear ODE's representable as
the compatibility condition of Lax pairs.
Vectors $\varphi$ and $\chi$ cannot be regarded as the solutions of systems
(\ref{1},\ref{2}), since any transformation of the second equations of these
Lax pairs ought to have no the kernel.
The basic idea of the generalization of the DT technique for ODE's is to put
the points, in which the DT of the first equations of corresponding Lax pairs
has the kernel, into the singular points of the second equations of Lax pairs.
For this aim we will assume in the next section that $\varphi$ and $\chi$ are
the solutions of systems (\ref{1},\ref{2}) in the points $\lambda=\mu$ and
$\lambda=\nu$.
Then we will consider the limits in the formulas of DT's, tending the points $\mu$ and
$\nu$, in which the kernel exists, to singular point $\lambda=0$.

\section{Transformations of ODE's}
To fulfill the procedure suggested in the previous section we need the explicit
expressions of the asymptotic expansions of solutions of Lax pairs
(\ref{1},\ref{2}) with coefficients given by Eqs.(\ref{5},\ref{6}) or
Eqs.(\ref{8},\ref{9}) at singular point neighborhood.

If $\alpha$ is unequal to half--integer, the asymptotic expansion of
the Lax pairs solutions of $\rm P_{II}$ at neighborhood of point $\lambda=0$ have the
form
\begin{equation}
\Psi = \Psi_0 ( E + \lambda \rho + o(\lambda) ) \Lambda ,
\label{18}
\end{equation}
\begin{equation}
\Xi = \Lambda^{-1} ( E - \lambda \rho + o(\lambda) ) \Psi_0^{-1} .
\label{19}
\end{equation}
Here we use notations
\begin{equation}
\Psi_0 =
\left(
\begin{array}{cc}
d& d^{-1} \\
d& -d^{-1}
\end{array}
\right) , \,\,\,
\Lambda =
\left(
\begin{array}{cc}
\lambda^\alpha& 0\\
0& \lambda^{-\alpha}
\end{array}
\right) , \,\,\,
\rho =
\left(
\begin{array}{cc}
0 & \Delta_1 \\
\Delta_2 & 0
\end{array}
\right) ,
\label{20}
\end{equation}
$$
\Delta_1 = \displaystyle \frac{(2 u_x + 2 u^2 + x) d^{\,-2}}{2\alpha - 1} , \,\,\,
\Delta_2 = \displaystyle \frac{(2 u_x - 2 u^2 - x) d^{\,2}}{2\alpha + 1} , \,\,\,
d = \mbox{exp} \left( \int u \,dx \right) .
$$
The expansions in series of solutions of Lax pairs contain the logarithmic
terms in the case of half--integer $\alpha$.

The asymptotic expansions of the solutions of Lax pairs of Eq.(\ref{7})
at neighborhood of point $\lambda=0$ are given by Eqs.(\ref{18},\ref{19}),
in which the coefficients are defined as follows:
\begin{equation}
\Psi_0 =
\left(
\begin{array}{cc}
c_1 & c_2 \\
\displaystyle \frac{\alpha - z}{w z} c_1 &
\displaystyle - \frac{c_2}{w}
\end{array}
\right) , \,\,\,
\Lambda =
\left(
\begin{array}{cc}
\mbox{e}^{\left( \displaystyle - \frac{\alpha}{2 \lambda} +
\frac{\gamma}{\alpha} \ln \lambda \right)} & 0\\
0& \mbox{e}^{\left( \displaystyle \frac{\alpha}{2 \lambda} -
\frac{\gamma}{\alpha} \ln \lambda \right)}
\end{array}
\right) ,
\label{21}
\end{equation}
where functions $c_1$ and $c_2$ solve equations:
$
c_{1,x} = \frac{\displaystyle \alpha - z}{\displaystyle x}  u c_1 , \,\,\,
c_{2,x} = - \frac{\displaystyle z}{\displaystyle x}  u c_2
$.

These asymptotic expansions allow us to apply formulas of the DT's for
deriving the transformations of ODE's.
Having the reduction constraint on potential $P^{(0)}$, we start from BDT
(\ref{16},\ref{17}) in the case of $\rm P_{II}$.
To satisfy condition $\mbox{Sp} \, Q = 0$ under performing the transformation
we put $\sigma = 1$.
Substituting
\begin{equation}
\left. \varphi = \Psi \, \right|_{\lambda = \mu} \, {a_1 \choose a_2} \, ,
\label{22}
\end{equation}
\begin{equation}
\left. \chi = (b_1\,,\,\,b_2) \,\,\, \Xi \, \right|_{\lambda = \nu}
\label{23}
\end{equation}
($a_1$, $a_2$, $b_1$, $b_2$ are constants) and considering the limits
$\mu \to 0$ and $\nu \to 0$ in the BDT formulas, taking into account
Eqs.(\ref{18},\ref{19},\ref{20}), we obtain two transformations of Lax
pairs solutions and corresponding well--known transformations of $\rm P_{II}$
(e.g., see \cite{15}):
$$
\left\{
\begin{array}{l}
\displaystyle \tilde \Psi = \Psi +
\ln_x (\Delta_1) \, ( \sigma_3 + i \sigma_2) \Psi /(2\lambda) \\
\displaystyle \tilde \Xi = \Xi -
\ln_x (\Delta_1) \, ( \sigma_3 + i \sigma_2) \Xi /(2\lambda) \\
\tilde u = u + \ln_x \Delta_1 \\
\tilde \alpha = 1 - \alpha
\end{array}
\right. , \,\,\,
\left\{
\begin{array}{l}
\displaystyle \tilde \Psi = \Psi +
\ln_x (\Delta_2) \, ( \sigma_3 - i \sigma_2) \Psi /(2\lambda) \\
\displaystyle \tilde \Xi = \Xi -
\ln_x (\Delta_2) \, ( \sigma_3 - i \sigma_2) \Xi /(2\lambda) \\
\tilde u = u - \ln_x \Delta_2 \\
\tilde \alpha = - 1 - \alpha \\
\end{array}
\right. .
$$

The formulas of transformations coincide with presented above or follow from
them as limits in the case of half--integer $\alpha$.

The transformations of Eq.(\ref{7}) are constructed similarly to that of
$\rm P_{II}$.
Substituting the asymptotic expansions (\ref{18},\ref{21}) in
formulas of EDT's (\ref{12},\ref{14}) accordingly to
Eq.(\ref{22}) and considering limit $\mu \to 0$, we obtain four
transformations of solutions and coefficients of Lax pairs:
$$
\left\{
\begin{array}{l}
\displaystyle \tilde \Psi = \left(
\begin{array}{cc}
\lambda + \displaystyle \frac{u_1}{2}_{\mathstrut}  \varepsilon_1^{\pm} &
- \displaystyle \frac{u_1}{2} \\
- \varepsilon_1^{\pm} & 1
\end{array}
\right)_{\mathstrut} \frac{\Psi}{\sqrt{\lambda}} \,\, , \,\,\,
\displaystyle \tilde \Xi = \frac{\Xi}{\sqrt{\lambda}} \left(
\begin{array}{cc}
1 & \displaystyle \frac{u_1}{2}_{\mathstrut} \\
\varepsilon_1^{\pm}  &
\displaystyle \lambda + \frac{u_1}{2} \varepsilon_1^{\pm}
\end{array}
\right) , \\
\displaystyle \tilde \beta = \beta + \frac{1}{2} , \,\,\,
\displaystyle \tilde v_1 = A_1^{\pm} +
\frac{v_1}{2x}_{\mathstrut} , \,\,\,
\displaystyle \tilde v_2 = 2 x \varepsilon_1^{\pm} , \,\,\,
\displaystyle \tilde w_1 = \frac{v_1}{2 x} \left( 2 q -
\frac{v_1 v_2}{2 x} + \right. \\
\displaystyle \left. + w_1 \varepsilon_1^{\pm} +
A_1^{\pm} \varepsilon_1^{\pm} \vphantom{\frac{v_1}{2}} \right)_{\mathstrut} , \,\,\,
\displaystyle \tilde w_2 = v_2 - 2 \beta \varepsilon_1^{\pm} -
v_1 \varepsilon_1^{\pm\,2} , \,\,\,
\displaystyle \tilde q = q - \frac{v_1 v_2}{2 x} + A_1^{\pm}
\varepsilon_1^{\pm} ,
\end{array}
\right.
$$
$$
\left\{
\begin{array}{l}
\displaystyle \tilde \Psi = \left(
\begin{array}{cc}
1 & - \varepsilon_2^{\pm} \\
\displaystyle \frac{u_2}{2}_{\mathstrut} &
\displaystyle \lambda - \frac{u_2}{2} \varepsilon_2^{\pm}
\end{array}
\right)_{\mathstrut} \frac{\Psi}{\sqrt{\lambda}} \,\, , \,\,\,
\displaystyle \tilde \Xi = \frac{\Xi}{\sqrt{\lambda}} \left(
\begin{array}{cc}
\lambda - \displaystyle \frac{u_2}{2}_{\mathstrut} \varepsilon_2^{\pm} &
\varepsilon_2^{\pm} \\
- \displaystyle \frac{u_2}{2}  & 1
\end{array}
\right) , \\
\displaystyle \tilde \beta = \beta - \frac{1}{2} , \,\,\,
\displaystyle \tilde v_1 = - 2 x \varepsilon_2^{\pm} , \,\,\,
\displaystyle \tilde v_2 = A_2^{\pm} - \frac{v_2}{2x}_{\mathstrut} , \,\,\,
\displaystyle \tilde w_1 = v_1 + 2 \beta \varepsilon_2^{\pm} -
v_2 \varepsilon_2^{\pm\,2} , \\
\displaystyle \tilde w_2 = - \frac{v_2}{2 x} \left( - 2 q +
\frac{v_1 v_2}{2 x} + w_2 \varepsilon_2^{\pm} +
A_2^{\pm} \varepsilon_2^{\pm} \vphantom{\frac{v_1}{2}} \right)_{\mathstrut} , \,\,\,
\displaystyle \tilde q = q - \frac{v_1 v_2}{2 x} - A_2^{\pm}
\varepsilon_2^{\pm} ,
\end{array}
\right.
$$
where
$
\displaystyle \varepsilon_1^{\pm} = \frac{1}{\varepsilon_2^{\pm}} = -
\frac{2 q \pm \alpha}{2 w_1} , \,\,\,
A_1^{\pm} = w_1 + \frac{\beta}{x} v_1 + \frac{v_1^2}{2 x} \varepsilon_1^{\pm} ,
\,\,\,
A_2^{\pm} = w_2 + \frac{\beta}{x} v_2 - \frac{v_2^2}{2 x}
\varepsilon_2^{\pm} .
$
(To keep condition $\mbox{Sp} \, Q = 0$ we assume $\sigma = \lambda^{-1/2}$ in
the formulas of EDT's.)

These transformations lead to the following ones of solutions and
parameters of Eq.(\ref{7})
$
\{u,\alpha, \beta, \gamma\} \stackrel{\displaystyle T_i}{\longrightarrow}
\{\tilde u, \tilde \alpha, \tilde \beta, \tilde \gamma\}
$
$(i = 1,...,4)$:
$$
\displaystyle T_1 : \,\,\,
\tilde u = - \frac{2x ( x u_x + \alpha u^2 - 2 x - 2 ( \beta + 1 ) u)}
{u ( \alpha (x u_x + \alpha u^2 - 2 x ) - 2 \gamma u)} , \,\,\,
\tilde \alpha = \alpha , \,\,\, \tilde \beta = \beta + \frac{1}{2} , \,\,\,
\tilde \gamma = \gamma - \frac{\alpha}{2} ;
$$
$$
\displaystyle T_2: \,\,\,
\tilde u = \frac{2x ( x u_x - \alpha u^2 - 2 x - 2 ( \beta + 1 ) u)}
{u ( \alpha (x u_x - \alpha u^2 - 2 x ) + 2 \gamma u)} , \,\,\,
\tilde \alpha = \alpha , \,\,\, \tilde \beta = \beta + \frac{1}{2} , \,\,\,
\tilde \gamma = \gamma + \frac{\alpha}{2} ;
$$
$$
\displaystyle T_3 : \,\,\,
\tilde u = \frac{2x ( x u_x + \alpha u^2 + 2 x + 2 \beta u)}
{u ( \alpha (x u_x + \alpha u^2 + 2 x ) - 2 \gamma u)} , \,\,\,
\tilde \alpha = \alpha , \,\,\, \tilde \beta = \beta - \frac{1}{2} , \,\,\,
\tilde \gamma = \gamma - \frac{\alpha}{2} ;
$$
$$
\displaystyle T_4 : \,\,\,
\tilde u = - \frac{2x ( x u_x - \alpha u^2 + 2 x + 2 \beta u)}
{u ( \alpha (x u_x - \alpha u^2 + 2 x ) + 2 \gamma u)} , \,\,\,
\tilde \alpha = \alpha , \,\,\, \tilde \beta = \beta - \frac{1}{2} , \,\,\,
\tilde \gamma = \gamma + \frac{\alpha}{2} .
$$

The EDT's of dual pair give the same transformations.
The transformations obtained commute and transformations $T_4$ and $T_3$ are inverse
to $T_1$ and $T_2$ respectively.
One can note that Eq.(\ref{7}) is invariant under changing
$\alpha \to - \alpha$.
This symmetry connects transformations $T_1$ and $T_2$, $T_3$ and $T_4$.
It should be stressed that a form of these transformations differs from one
considered in \cite{15}.

If $\gamma = \pm \alpha \beta$ or $\gamma = \pm \alpha (\beta + 1)$,
particular solutions of Eq.(\ref{7}) satisfy the Riccati equations.
In these cases we obtain, performing the transformations of Eq.(\ref{7}),
which keep this condition, two transformations
$$
\left\{
\begin{array}{l}
\displaystyle \tilde \psi = \sqrt{x} \left(\psi_x - \frac{\beta+1}{x} \psi
\right)_{\mathstrut} \\
\displaystyle \tilde \beta = \beta + \frac{1}{2}
\end{array}
\right. , \,\,\,
\left\{
\begin{array}{l}
\displaystyle \tilde \psi = \sqrt{x} \left(\psi_x + \frac{\beta}{x} \psi
\right)_{\mathstrut} \\
\displaystyle \tilde \beta = \beta - \frac{1}{2}
\end{array}
\right.
$$
of the linearization of mentioned Riccati equations:
$
\displaystyle \psi_{xx} - \left(\frac{2\alpha}{x} + \frac{\beta(\beta+1)}{x^2}
\right) \psi = 0 .
$

So, the modification of the DT technique presented here allows us to obtain
the transformations of solutions of the nonlinear ODE's admitting the
compatibility condition representation.
At the same time we derive the Schlesinger transformations of associated Lax
pairs.
This approach can be applied for finding the transformations of nonlinear
systems of ODE's and their Lax pairs that have different types of the
singularities and the reduction constraints of the coefficients.

\end{document}